\documentclass[%
 reprint,
]{revtex4-2}
\usepackage{amsmath} 
\usepackage{graphicx}
\usepackage{dcolumn}
\usepackage{bm}
\usepackage{amssymb}
\usepackage{hyperref}
\usepackage{xcolor}
\usepackage [english]{babel}
\usepackage [autostyle, english = american]{csquotes}
\MakeOuterQuote{"}

\begin{document}

\preprint{APS/123-QED}

\title{Coherently Driven Quantum Harmonic Oscillator Battery}

\author{Kuldeep Gangwar}
\author{Anirban Pathak}%
\affiliation{Department of Physics and Materials Science \& Engineering, Jaypee Institute of Information Technology, A-10, Sector-62, Noida, UP 201309, India}
 \email{anirban.pathak@jiit.ac.in}
\date{\today}

\begin{abstract}
 Quantum harmonic oscillator (QHO) battery models have been studied with significant importance in the recent past because these batteries are experimentally realizable and have high ergotropy and capacity to store more than one quanta of energy. QHO battery models are reinvestigated here to answer a set of fundamental questions: Do such models have any benefit? Is unbounded charging possible? Does the use of a catalyst system enhance the energy transfer to quantum batteries? These questions are answered both numerically and analytically by considering a model that allows a laser to shine on a QHO charger that interacts with a QHO battery. In contrast to some of the existing works, the obtained answers are mostly negative. Specifically, in the present work, the laser frequency is tuned with the frequency of the global charger-battery system, which is affected by the interaction between QHOs. It is reported that for a fixed laser field amplitude \textit{F}, the battery can store more energy when tuned with the frequency of the global charger-battery system compared to energy stored by tuning the laser frequency with local frequencies of the charger and battery. The charging process of the open QHO, which is a simplified model, and the self-discharging (dissipation) process after switching off the laser field are also investigated to reveal that the charging process of QHO in the simplified model is faster than the charging process of the catalytic (non-catalytic) battery. Further, it's observed that the self-discharging process is almost two times faster than the charging process which makes such models unstable against interaction with the environment. 
\end{abstract}
\maketitle


\section{\label{sec:level1}Introduction}
Quantum thermodynamics is the study of two independent theories: thermodynamics and quantum mechanics. Specifically, quantum thermodynamics explores how the laws of thermodynamics emerge from the underlying framework of quantum mechanics \cite{VA16}. Moreover, quantum thermodynamics explores how traditional thermodynamic principles apply in the quantum realm, where quantum effects like quantum fluctuations, quantum coherence, and entanglement play a significant role. With the advent of the second quantum revolution \cite{DM03}, we are in a situation to build several extremely small quantum devices. There is a need to supply energy on demand to such devices and to store that energy in systems of similar dimensions. This specific need is addressed by a quantum battery- a device that can temporarily store energy and supply that (as and when needed) to another device just like a traditional battery, but in contrast to the conventional batteries it exploits the nonclassical features of the quantum world \cite{AKM+19,CPF+17}. As we strive to create smaller quantum devices for emerging technologies, such as special-purpose quantum computing devices and quantum sensors, the need for storing energy in small devices by exploiting nonclassical features of quantum mechanics and using the stored energy on demand, has been considerably increased. To address this need, the idea of the quantum battery was introduced. A quantum battery is a quantum device that stores energy temporarily, and this energy can be extracted as work to perform a task. The performance of the battery is characterized primarily by three parameters: (1) the amount of energy that this battery can store (2) the amount of energy that can be extracted as work, and (3) charging time. However, we think that the inclusion of one more parameter called self-discharging time would provide completeness to this list and allow one to analyze the stability of a battery. Before we elaborate on this point, it will be apt to summarize the existing works on quantum batteries related to the present work.

In 2013, Alicki and Fannes \cite{AF13} published a seminal work on quantum battery, where they showed that entangling unitary controls acting globally extract more work, in general, than local unitary operation on each cell. Another seminal work was put forward by Hovhannisyan et al. \cite{HK+13} in the same year, where they demonstrated that maximal work extraction is not connected to entanglement, but the power at which work is extracted is connected to the entanglement for $N$ non-interacting particles. These early works on quantum batteries and their potential applications developed an enhanced interest among the researchers and that has led to several charger-battery models of quantum batteries. These charger-battery models for quantum batteries can be classified into three types of models: (1) Qubit-Qubit (2) quantum harmonic oscillator (QHO)-Qubit and (3) QHO-QHO models. Among these classes, in Qubit-Qubit model, the battery can store only one quantum of energy for a qubit and does not have any collective quantum advantage \cite{AKM+19}. Therefore, this model does not get much attention. In QHO-Qubit model, the battery charges fastest \cite{AF+18}. However, when $N$ non-interacting qubits interact with the single QHO, charging time is observed to be reduced by a factor of $\sqrt{N}$. This battery model is known as the Dicke quantum battery due to the similarity of the physical system with the well-known Dicke model \cite{D-54}. In Dicke quantum battery the quantum advantage \cite{CPF+17} originates from the collective behaviour of the  qubits \cite{AKM+19,ZB-23,DCA+18,CCS+20,RAR+20,CLL+20,ZYF+19,CYD+21,S-21,LLM+18,SS-21,PHW+21,JS+20,MR}. In fact, the Dicke quantum battery has received a huge attention due to a clear quantum advantage and the fact that it charges almost instantaneously. This almost instantaneous charging phenomenon is referred to as the extensive superabsorption of energy and that's what makes the Dicke quantum battery interesting. Further, the Dicke quantum battery has been experimentally realized \cite{QMK+22}. Since coherence phenomena are symmetric in time, we expect superradiance when this battery discharges. Such batteries have potential applications, as supercapacitors, where we need high power for short-time. However, the ergotropy of the Dicke quantum battery has not been discussed so far which is also an important quality factor of an efficient quantum battery.

On the other hand, we have another class: QHO-QHO model, such battery models are called QHO battery in short where the charger and battery both are considered to be QHOs. A QHO battery can store an arbitrary number of quanta in a single QHO (subjected to experimental limitations). In contrast to this battery, to store an arbitrary number of quanta in a Dicke battery, an arbitrary number of qubits will be needed, and achieving coherence for a long time for such a big system would be very difficult. Besides this, QHO battery has high ergotropy and these batteries are experimentally realizable using the available superconducting technology (cf. Section IV of \cite{RA-23} and references therein). However, the charging of the QHO battery is slower than the Dicke quantum battery, but such QHO battery models have potential applications where we need power for a relatively longer duration. There is another way to classify quantum batteries based on whether a "catalyst", specifically a QHO or a qubit is incorporated between the charger and the battery. If such a catalyst is used (not used) in a battery model then the battery model is referred to as the catalytic (non-catalytic) model \cite{RA-23}. Here, it may be noted that the catalyst serves as a mediator of energy transfer.

Keeping the above in mind, in this paper, we will investigate charger-mediated energy transfer for QHO batteries via an open system approach considering both non-catalytic and catalytic models \cite{RA-23}. In what follows, we will discuss the charging and self-discharging time of catalytic, and non-catalytic models and compare them with the corresponding quantities for a simplified battery model where no separate charger will be considered and the environment will be treated as part of the charging system.

The rest of the paper is organised as follows. In \hyperref[sec:level2]{Sec-II}, we briefly describe the QHO battery model studied in the present work and the approach adopted for this investigation. In \hyperref[sec:level3]{Sec-III}, we investigate both analytically and numerically the energy, and ergotropy behaviour with time in the off-resonance battery charging setup, i.e., when laser frequency is out of tune with the global frequency of charger-battery system. Specifically, in the numnerical part, we solve Gorini-Kossakowski-Sudarshan-
Lindblad (GKSL) master equation numerically by using QuTiP packages \cite{JNN-13}.  Subsequently, in \hyperref[sec:level4]{Sec-IV}, we perform a similar numerical investigation on the energy behaviour of the battery and charger with time by setting laser frequency on resonance with the global frequency of the charger-battery system. In the next section (i.e., in \hyperref[sec:level5]{Sec-V}) an analytic solution for the on-resonance case is provided, and the solution is used to compare the analytic results with the numerical results produced by QuTiP and summarized in the previous section. In \hyperref[sec:level6]{Sec-VI}, we present a simplified form of QHO battery model and investigate the charging and self-discharging process. Finally, the paper is concluded in \hyperref[sec:level7]{Sec-VII}.

\section{\label{sec:level2}Model and Approach}
\begin{figure}[th]
    \centering
    \includegraphics[width=\linewidth]{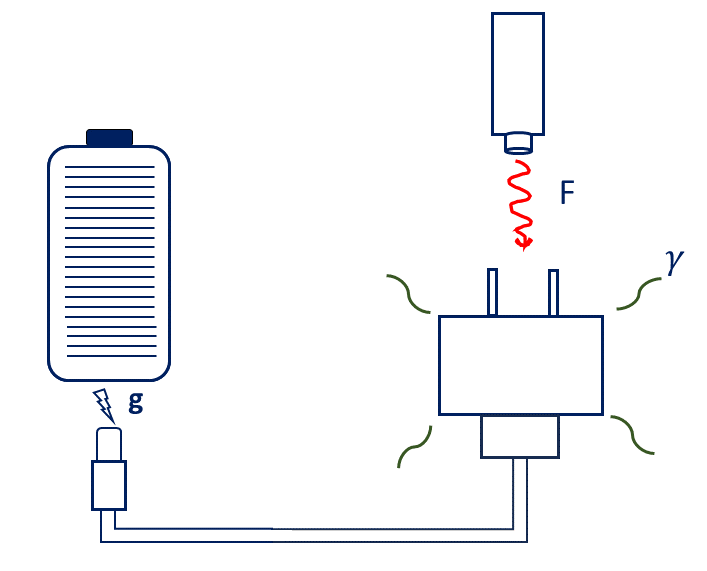}
    \caption{(Color Online) A QHO as charger (bottom right), which interacts with another QHO treated as a battery (left) and driven by coherent driving force (top right). Green curved lines indicate energy dissipation of charger to the environment.}
    \label{fig:fig-01}
\end{figure}

In this section, we present the model of the charger-battery system studied in the present work. Here, we consider a QHO as a charger, which interacts with another QHO treated as a battery. The interaction between the charger and the battery takes place for the time period $(0,\tau)$, where $\tau$ is the charging time. The charger also interacts with the environment and dissipates energy. To charge the battery, the charger is driven by a laser field of amplitude $F$. The charger-battery system studied here is schematically shown in \hyperref[fig:fig-01]{Figure-1} and can be mathematically described by the total Hamiltonian of the charger-battery \cite{FA+19} system expressed as
\begin{align}
{\cal H} &= \omega_o a^\dagger a + \omega_o b^\dagger b + g(ab^\dagger + a^\dagger b) \nonumber \\
&\quad+F(a e^{i\omega_ft}+  a^\dagger e^{-i\omega_ft}),\label{eq:1}
\end{align}
where $\omega_o$ is frequency of charger and battery, $\omega_f$ is laser frequency, $a^\dagger$ are annihilation and creation operators of charger (A), and $b$ and $b^\dagger$ are annihilation and creation operators of battery (B). Here \textit{g} is the coupling strength between charger and battery which is positive for time period $(0,\tau)$, and \textit{F} is the amplitude of the laser field. As the energy non-conserving terms will oscillate much faster compared to the energy conserving terms, we can neglect fast oscillating terms of charger-battery interaction under rotating wave approximation.
\par
For convenience, we can express the above Hamiltonian of the system in terms of supermode operators $C_{\pm}$ as

\begin{eqnarray}
{\cal H} &=& \omega_{+} C_{+}^\dagger C_{+}+ \frac{F}{\sqrt{2}}(C_{+} e^{i\omega_{+}t}+  C_{+}^\dagger e^{-i\omega_{+}t}) \nonumber \\
&+& \omega_{-} C_{-}^\dagger C_{-}+ \frac{F}{\sqrt{2}}(C_{-}e^{i\omega_{-}t}+ C_{-}^\dagger e^{-i\omega_{-}t}), \label{eq:2}
\end{eqnarray}
where
\begin{eqnarray}
C_{\pm}= \frac{1}{\sqrt{2}}(a\pm b),\label{eq:3}
\end{eqnarray}
and
\begin{eqnarray}
\omega_{\pm}=\omega_o\pm g. \label{eq:4}
\end{eqnarray} 
Here, $\omega_{\pm}$ is global frequency of charger-battery system. 
\par
In our model, it is assumed that both charger and battery are initially in the ground state. Since charger interacts with environment and dissipates energy to the environment, therefore the evolution of state of the system would be given by GKSL master equation \cite{Breuer:2002pc,SB}
\begin{align}
{\partial_t\rho_{AB} (t)}&= -i[{\cal H}, \rho_{AB}(t)] + \gamma (N(T)+1) D_a[\rho_{AB}]\nonumber \\
&\quad+\gamma N(T) D_a[\rho_{AB}],  \label{eq:5}   
\end{align}
where $\rho_{AB}(t)$ represents the state of charger \textit{A} and battery \textit{B}. Here, $D_a[\rho_{AB}]= a\rho_{AB} a^\dagger- \frac{1}{2}\left\{a^\dagger a,\rho_{AB}\right\}$ is dissipator and $\gamma$ is the dissipation constant. $N(T) = 1/(e^{\frac{\omega_e}{kT}}-1)$ is the average number of photons in environment mode of frequency $\omega_e$, \textit{k} is Boltzmann constant, and \textit{T} is temperature of the environment.
\\

To characterize the performance of battery we calculate energy expectation value of the battery using a parameter defined as follows \cite{AF+18}:
\begin{eqnarray}
E_{B} (t) = \omega_o \langle b^\dagger b\rangle (t).\label{eq:6}
\end{eqnarray}
Since not all the stored energy can be extracted as work, therefore we may define another parameter that is known as ergotropy \cite{TK+20,FB+20,B19}, the maximum energy that can be extracted from the battery as work by cyclic unitary operations. Ergotropy is defined as follows:
\begin{eqnarray}
{\cal W} (t) = E_{B} (t)- tr(\rho^p (t) b^\dagger b), \label{eq:7}
\end{eqnarray}
where $\rho^p$ is a passive state \cite{AF13,PW-78} of battery. A passive state is defined as the state from which energy cannot be extracted. Mathematically, we can express a bound on the second term of the above equation as  
\begin{eqnarray}
    tr(\rho^p H_B) \leq tr(U^\dagger \rho^p U H_B). \label{eq:8}
\end{eqnarray}
To find $\rho^p (t)$, we first trace out charger state $\rho_B=tr_A[\rho(t)]$, and then diagonalize  $\rho_B$ to obtain
\begin{eqnarray}
\rho^p= \sum_{i}\lambda_i\ |\epsilon_i\rangle \langle \epsilon_i|, \label{eq:9}
\end{eqnarray}
where $\lambda_i > \lambda_{i+1}$ are eigenvalues of $\rho_B$, $\epsilon_{i+1}>\epsilon_{i}$ are energy eigenvalues of battery Hamiltonian.\\
\section{\label{sec:level3}Off-resonance battery charging: Numerical and Analytical Solution}
In this section, we present off-resonance battery charging which has already been explored in \cite{FA+19,RA-23}. Off-resonance means laser frequency is out of tune with the global frequency of charger-battery system $\omega_\pm$. In this case, laser frequency is tuned with local frequencies of charger and battery by considering $\omega_f$ = $\omega_o$.
\begin{figure}[th]
    \centering
    \includegraphics[width=\linewidth]{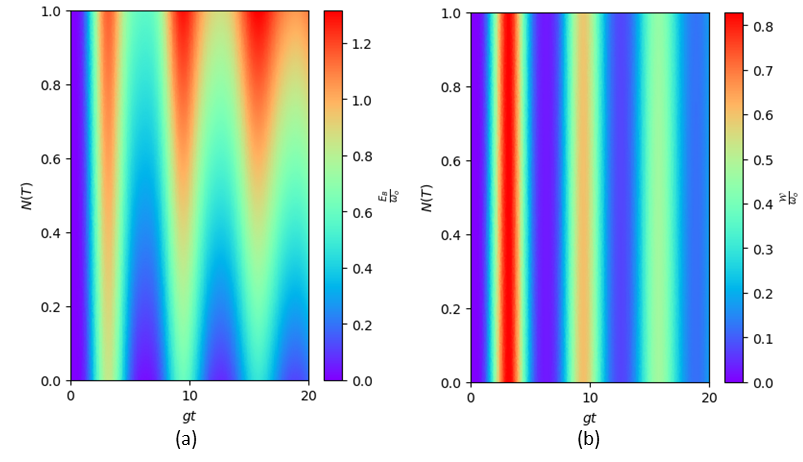}
    \caption{ (Color Online) We have plotted behavior of the stored energy $E_B$ in Fig.(a) and ergotropy ${\cal W}$ of the battery in Fig.(b) in units of $\omega_o$ ($\hbar=1$). In both cases, we have taken $g=0.2\omega_o$, $\textit{F}=0.1\omega_o$, $\gamma_A=0.05\omega_o$, $\gamma_B=0$, and $N(T)=0$ to $N(T)=1$.}
    \label{fig:fig-02}
\end{figure}
\\

In  \hyperref[fig:fig-02]{Figure-2(a)}, we plot numerical outcomes of the energy and the ergotropy considering laser frequency is tuned with local frequencies of charger and battery and is out of tune with global frequency, and charger is dissipative. In this case, we can see that the battery is storing only around 0.8 energy which is less than a quanta, and ergotropy is equal to stored energy, therefore all the stored energy can be extracted as work. At zero temperature, Markovian dynamics maps a coherent state into another coherent state \cite{KOSSAKOWSKI1972247}, therefore, ergotropy would be always equal to the stored energy in this particular case. From  \hyperref[fig:fig-02]{Figure-2(b)}, we can see that the battery is storing only around 1.2 energy, and ergotropy is constant. With temperature, battery energy is found to increase but ergotropy is found to remain constant because thermal energy is a passive form of radiation. For further explanation on bound of energy of battery when $N(T)=0$, we look into analytical expression \cite{FA+19}
\begin{align}
 E_B = \frac{\omega_o F^2}{g^2} \left[1 - e^{-\frac{\gamma t}{4}} \left(\cosh\left(\frac{\epsilon t}{4}\right) + \frac{\gamma}{\epsilon}\sinh\left(\frac{\epsilon t}{4}\right)\right)\right]^2, \label{eq:10}   
\end{align}
where $\epsilon=\sqrt{\gamma^2-(4g)^2}$. From Eq. (10), we can conclude that the energy of the battery is asymptotically ($t\rightarrow0$) bounded at the order of $\frac{\omega_{0}F^2}{g^2}$. The battery can store unbounded energy if $g\rightarrow0$. However, from \eqref{eq:10} we can see that charging time will diverge in that case.\par

Here, the battery is not charging efficiently because the laser is tuned with the local frequencies of the charger and battery, but it is out of tune with the global frequency $\omega_\pm$. Keeping this in mind, in the next section we will discuss the on-resonance battery charging and check whether that helps us in efficiently charging the battery.

\section{\label{sec:level4} On-Resonance Battery Charging: Numerical Solution}
In this section, we discuss on-resonance battery charging, where the laser frequency $\omega_f$ is tuned with global frequency $\omega_\pm$. We also discuss the impact of detuning between laser frequency and global frequency, $\Delta= \omega_\pm - \omega_f$. To efficiently charge a battery, we switch the laser frequency $\omega_f$ from $\omega$ to $\omega_+(\omega_-)$ to charge the supermode $C_+(C_-)$. 
\par
In the upper row of \hyperref[fig:fig-03]{Figure-3}, we have plotted the numerically obtained values for the energy of the battery and charger when the charger is non-dissipative. We investigate the behavior energy of the battery and charger by varying detuning $\Delta$. When $\Delta=0$, the energy of the battery keeps increasing \textit{boundlessly} with time, which can be understood from Eq.\eqref{eq:2}. From Eq.\eqref{eq:2}, it is clear that there are two independent forced QHOs when laser frequency is tuned with global frequency, effective charging takes place, and the laser keeps pumping energy into the battery. As a result, the energy of the battery increases boundlessly with time. But as we increase detuning, effective charging of the battery is compromised. As a result, the maximum value of energy decreases. \\
In the bottom row of \hyperref[fig:fig-03]{Figure-3}, we have presented the numerical outcomes concerning the energy of both the battery and charger under dissipative charger conditions. Our investigation involves an exploration of how the energy levels of the battery and charger evolve as we manipulate the detuning parameter $\Delta$. Contrary to the previous assertion made in Ref.  \cite{RA-23}, we observe that the battery's energy does not increase indefinitely. Instead, it reaches a maximum value of 16 and subsequently stabilizes. Furthermore, we notice that as the detuning parameter $\Delta$ increases, the maximum energy decreases, and the energy experiences intermittent oscillations before stabilizing. However, the noteworthy observation is that the energy accumulation does not stop.
\begin{widetext}

\begin{figure}[th]
    \centering
    \includegraphics[width=\linewidth]{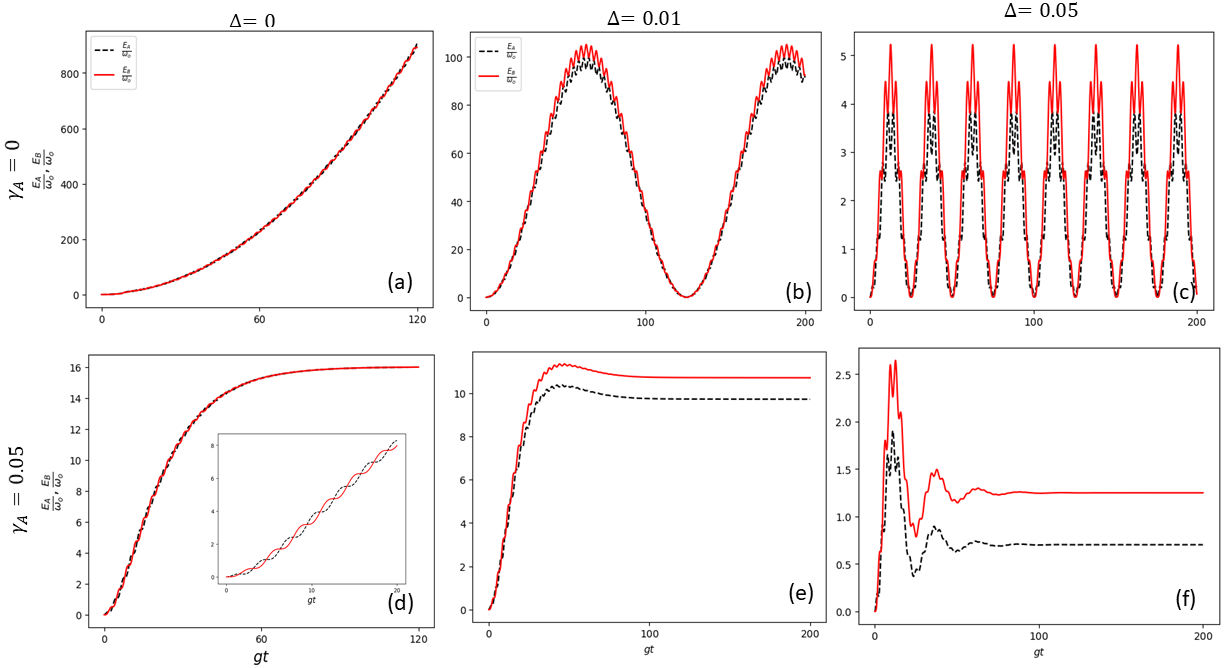}
    \caption{ (Color Online) Behavior of the charger energy $E_A$ and battery energy $E_B$ in units of $\omega_o$ ($\hbar=1$). In both cases, we have taken $g=0.2\omega_o$, $\textit{F}=0.1\omega_o$. Upper row $\gamma_A=0.$, $\gamma_B=0$ (a-c). Bottom row $\gamma_A=0.05\omega_o$, $\gamma_B=0$ (d-f). In the inset, we have shown that there are small oscillations of energy between charger and battery.}
    \label{fig:fig-03}
\end{figure}
\par
\end{widetext}

In the following section, we will present the model's analytical solution to provide a clearer understanding. While our numerical solutions have shown that detuning affects both the maximum energy and oscillation, the precise manner in which maximum energy and frequency depend on detuning remains ambiguous.

\section{\label{sec:level5} On-Resonance Battery Charging: Analytical Solution}
In this section, we aim to present an analytical solution to study the on-resonance model in great detail. To avoid time dependence complications, we first write the Hamiltonian Eq.\eqref{eq:1} in a rotating frame, which is rotating with angular velocity $\omega_f$. The Hamiltonian in the rotating frame is
\begin{align}
{\cal H_R}& =  \delta (a^\dagger a +  b^\dagger b) + g(ab^\dagger + a^\dagger b) \nonumber \\
&\quad + F(a +  a^\dagger),\label{eq:11}
\end{align}
where $\delta=\omega_o-\omega_f$. Detailed solution of ${\cal H_R}$ can be found in Appendix A. Now, we put ${\cal H_R}$ described in Eq.\eqref{eq:11} into GKSL master equation Eq.\eqref{eq:5} to obtain
\begin{align}
{\partial_t\rho_{AB} (t)}& = -i[{\cal H_R}, \rho_{AB}(t)] \nonumber \\
&\quad+ \frac{\gamma_A}{2}[2a\rho_{AB} (t)a^\dagger -a^\dagger a\rho_{AB} (t)-\rho_{AB} (t)a^\dagger a] ]\label{eq:12}.
\end{align}
Subsequently, solving Eq.\eqref{eq:12} for the first moments, we obtain the following relations needed for the present work.   

\begin{eqnarray}
\langle\dot{a}\rangle &= -i\delta\langle a \rangle - ig\langle b \rangle - \frac{\gamma_A}{2}\langle a \rangle - iF, \label{eq:13}
\end{eqnarray}
\begin{eqnarray}
\langle\dot{b}\rangle &= -i\delta\langle b \rangle - ig\langle a \rangle, \label{eq:14}
\end{eqnarray}
where we have used the fact that $\langle\dot{x}\rangle=Tr[x\dot{\rho}(t)]$ indicates expectation value of any operator $\dot{x}$. After solving these coupled differential equations (i.e., Eqs. \eqref{eq:13} and \eqref{eq:14}), we obtain $\langle b(t) \rangle$ and $\langle b^\dagger (t) \rangle$. We use mean-field approximation to get the energy expression of the battery. The mean-field approximation is precise when the number of excited particles in the battery is large enough.
Under mean-field approximation, we can write.
\begin{eqnarray}
    E_B =\omega_o \langle b^\dagger(t) b(t)\rangle\approx \omega_o\langle b^\dagger(t)\rangle \langle b(t)\rangle. \label{eq:15}
\end{eqnarray}
\begin{widetext}
Now denoting detuning as $\Delta=\delta - g=\omega_{-}-\omega_f$ and  assuming $\gamma_A^2 \approx0$ we can obtain an expression for $E_B$ as follows
\begin{align}
E_B(t) \approx & \frac{\omega_o F^2}{\{\Delta^2(\Delta+2g)^2+\frac{\gamma_A^2 (\Delta+g)^2}{4}\}}[g^2+g^2 e^{-\frac{\gamma_A t}{2}}-g e^{-\frac{\gamma_A t}{4}}\{\frac{\gamma_A}{2} \sin(gt) \cos[(\Delta+g) t]+(\Delta+2g)\cos(\Delta t) \nonumber \\
& -\Delta \cos[(\Delta+2g )t]\}-\frac{e^{-\frac{\gamma_A t}{2}}}{8}\{4\Delta(\Delta+2g)]\cos(2gt)-2\gamma_A (\Delta+g) \sin(2gt) \}], \label{eq:16a}
\end{align}
where analytic expressions of $\langle b (t)\rangle$ and $\langle b^\dagger(t) \rangle$ obtained by solving Eqs. \eqref{eq:13} and \eqref{eq:14} are substituted in Eq. \eqref{eq:15}. If we now consider that the charger does not interact with the environment $\gamma_A=0$, then the above analytic expression for $E_B(t)$ would reduce to 
\begin{eqnarray}
E_B(t) \approx \omega_o F^2 \left[\frac{2g^2}{\Delta^2(\Delta+2g)^2}-\frac{g\cos(\Delta t)}{\Delta^2(\Delta+2g)}+\frac{g\cos[(\Delta+2g )t]}{\Delta(\Delta+2g)^2}-\frac{\cos(2gt)}{2\Delta(\Delta+2g)}\right]. \label{eq:17a}
\end{eqnarray}
For small detunnig $2g>>\Delta$, we can further simplify Eq.\eqref{eq:17a} to yield
\begin{align}
    E_B(t) \approx \frac{\omega_o F^2}{\Delta^2} \sin^2\left(\frac{\Delta t}{2}\right) \label{eq:17b}
\end{align}
\begin{figure}[th]
    \centering
    \includegraphics[width=\linewidth]{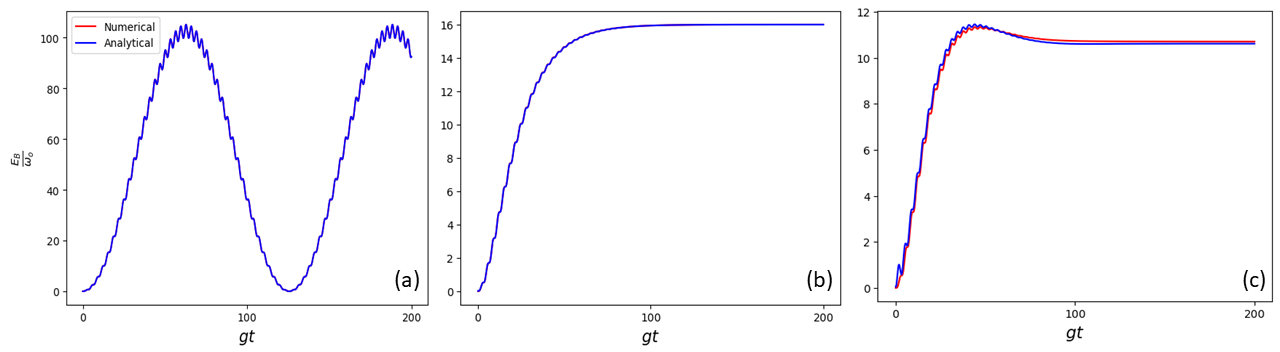}
    \caption{(Color Online) Behavior of the battery energy $E_B$ in units of $\omega_o$ ($\hbar=1$). (a) $\gamma_B=0$ and $\Delta=0.01$ (b) $\gamma_B=0.05\omega_o$ and $\Delta=0$ (c)$\gamma_B=0.05\omega_o$ and $\Delta=0.01$. In all cases, we have taken $g=0.2\omega_o$, $\textit{F}=0.1\omega_o$.}
    \label{fig:fig-04}
\end{figure}
\end{widetext}

A detailed explanation of Eq. \eqref{eq:16a} and Eq. \eqref{eq:17a} can be found in Appendix B. To show the accuracy of the mean-field approximation, we have plotted, \hyperref[fig:fig-04]{Figure-4}, numerical and analytical, Eq. \eqref{eq:16a} and Eq. \eqref{eq:17a}, results together. The mean-field approximation is so accurate when the number of excited particles is large that the numerical and analytical curves are observed to almost overlap with each other.
\\

From Eq. \eqref{eq:17a} and Eq. \eqref{eq:17b}, we can see that bigger amplitude oscillations in a non-dissipative charger are related to detuning parameter $\Delta$ and primarily caused by the term $\frac{g\cos(\Delta t)}{\Delta^2(\Delta+2g)}$. Further, a closer look into  \hyperref[fig:fig-04]{Figure-4(a)} and \hyperref[fig:fig-03]{Figure-03} upper row reveals that the amplitude of the oscillations, i.e., maximum energy, is roughly proportional to $1/\Delta^2$. The period of these oscillations is also observed to be proportional to $1/\Delta$, which are consistent with the numerical results \hyperref[fig:fig-03]{Figure-3} upper row.
\\

Now, we may discuss the more realistic case, when the charger is dissipative. From Eq.\eqref{eq:16a}, (also see \hyperref[fig:fig-04]{Figure-4(b, c)}), it is clear that the small oscillations in energy die out and energy becomes constant after some time because of dissipation of energy and maximum of energy of battery depends on detuning $\Delta$ and $\gamma_A$. It was reported in Refs. \cite{RA-23} that the battery's energy boundlessly increases with time and oscillates with period $1/\Delta$. However, we do not observe such behavior in the case of a dissipative charger.
\\

We wish to compare the battery energy behavior of the results depicted in \hyperref[fig:fig-04]{Figure-4} with those observed in the catalytic model \cite{RA-23}. The key distinction is observed to lie in the fact that, in the catalytic model, there is no need to measure $g$ for laser frequency calibration. Given that the charging processes are identical in both catalytic and non-catalytic models, our forthcoming comparison between the non-catalytic model and a simple model mirrors the comparison between the catalytic model and the same simple model.
\\

\section{\label{sec:level6} A Simplified Model of Quantum Battery}
To illustrate the fundamental aspects of QHO batteries, we may now introduce a simplified model of quantum battery. In this setup, we illuminate a QHO with a laser field of the same amplitude. The QHO  interacts with the same environment as in the previous model where the charger was interacting with the environment. We are not considering charger as a separate system here, instead, we are considering the environment as the part of "charging process". We consider the following Hamiltonian:
\begin{align}
{\cal H_R}& =  \delta a^\dagger a + F(a +  a^\dagger).\label{eq:18}
\end{align}
We derive analytical QHO energy expression for the above Hamiltonian under mean-field approximation as 
\begin{align}
    E_A\approx\frac{\omega_oF^2}{\frac{\gamma_A^2}{4}+\Delta^2}[e^{-\gamma_A t}-2e^{-\frac{\gamma_A t}{2}}\cos{\delta t}+1].\label{eq:19}
\end{align}
\begin{widetext}
    
\begin{figure}[th]
    \centering
    \includegraphics[width=\linewidth]{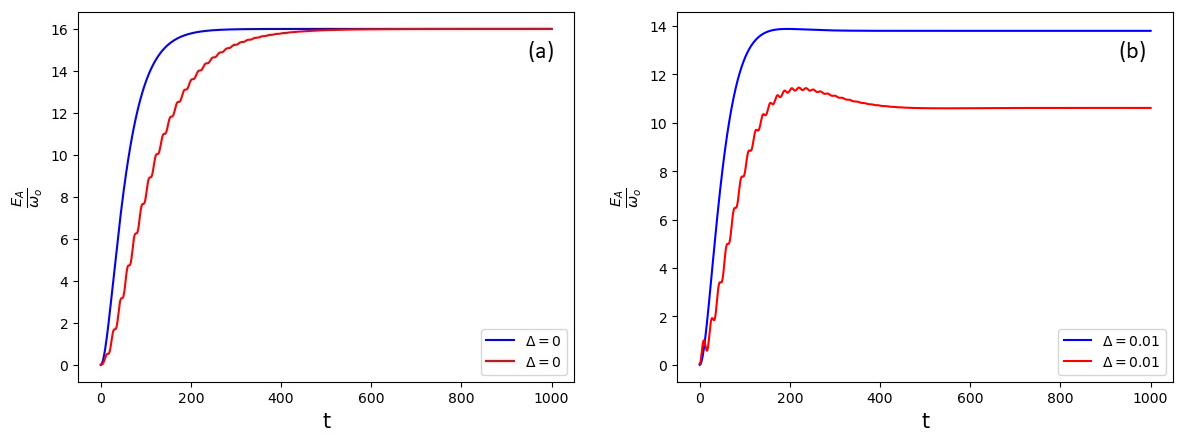}
    \caption{(Color Online) {\bf Charging:} Behavior of energy of the simple QHO battery and catalytic(non-catalytic) battery $E_A$ in units of $\omega_o$ $(\hbar=1)$. Fig(a) and Fig(b) correspond to different detunings. In Blue curves correspond to the simplified model and red curves correspond to the catalytic (non-catalytic) model. Amplitude of the laser field $F=0.1\omega_o$, dissipation constant for battery $\gamma_A=0.05\omega_o$.} 
    \label{fig:fig-05}
\end{figure}
\end{widetext}
Here, $\Delta=\delta=\omega_o-\omega_f$ is detuning between QHO frequency and laser field frequency. In \hyperref[fig:fig-05]{Figure-5}, we present a comparative plot of Eq.\eqref{eq:16a} and Eq.\eqref{eq:19} for various values of detuning ($\Delta$) to facilitate visual comparison between simple QHO battery and catalytic (non-catalytic) battery. From \hyperref[fig:fig-05]{Figure-5}, we can notice that the charging process of a simple QHO battery is faster compared to the charging process of a catalytic (non-catalytic) battery. For detuning $\Delta=0.01$,  the energy of the simple QHO battery is also greater than the catalytic (non-catalytic) battery.\\
It is known that the charged battery leaks out energy to the environment. This phenomenon is referred to as the self-discharging of battery. It is unavoidable, but this leakage process of energy should not be faster than the charging process for a stable battery. To visualize and compare the stability of the various battery models discussed here,  we have allowed the battery to interact with the environment (thermal bath) after being charged (i.e., after the battery is charged). The energy of both catalytic (non-catalytic) battery and QHO are found to decay exponentially. \\
A bit of calculation reveals that the self-discharging of simple QHO battery energy would follow the following relation
\begin{align}
    E_A\approx\frac{\omega_oF^2 e^{-\gamma_A t}}{\frac{\gamma_A^2}{4}+\Delta^2}[e^{-\gamma_A \tau}-2e^{-\frac{\gamma_A \tau}{2}}\cos{\delta \tau}+1].\label{eq:20}
\end{align}
\begin{figure}[th]
    \centering
    \includegraphics[width=\linewidth]{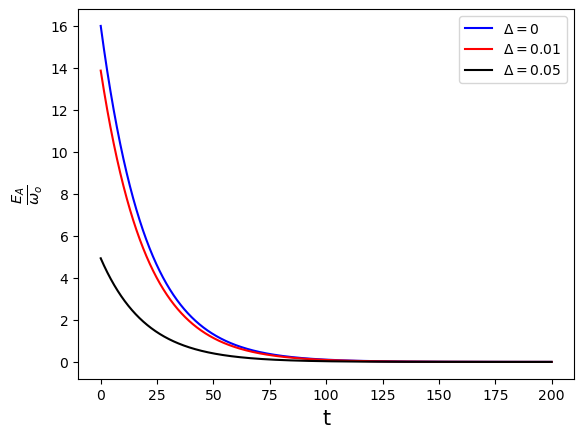}
    \caption{(Color Online) {\bf Self-discharging:} Behaviour of energy of the simple QHO battery $E_A$ in units of $\omega_o$ $(\hbar=1)$ with time for different detuning $\Delta$. Amplitude of the laser field $F=0$, dissipation constant $\gamma_A=0.05\omega_o$.}
    \label{fig:fig-06}
\end{figure}
Here, $\tau$ is the charging time of the QHO. After time t= $\tau$ (i.e., when the battery is charged), we switch off the laser. In \hyperref[fig:fig-06]{Figure-6}, we plot energy expression Eq.\eqref{eq:20} for different values of detuning $\Delta$. We can notice in \hyperref[fig:fig-05]{Figure-5} and \hyperref[fig:fig-06]{Figure-6}  that the self-discharging process is almost two times faster than the charging process. Consequently, the energy of a simple QHO battery is quite unstable against the interaction with the environment. Since the charging process of catalytic and non-catalytic batteries is slower than the simple QHO battery, these batteries are more unstable than the simple QHO battery.   \\

\section{\label{sec:level7}Conclusion}
In this work, we have investigated charger-mediated energy transfer for QHO quantum batteries via an open-system approach for a non-catalytic model. We have shown that instead of tuning laser frequency $\omega_f$ with local frequencies $\omega_o$ of the charger-battery system, we may tune laser frequency $\omega_f$ with global frequency $\omega_\pm$; in this way, the battery can store more energy. Further, in contrast to the previous study \cite{RA-23}, we have not observed any signature of boundless charging and oscillation of period $1/\Delta$, when the charger is dissipative. In fact, in contrast to \cite{RA-23}, it's established that catalysts are not essential for the efficient charging of QHO batteries. Finally, we have presented another simplified model of QHO batteries in which a QHO battery is found to be charged faster than catalytic and non-catalytic batteries. We have also pointed out that self-discharging of catalytic and non-catalytic batteries and QHO is faster than the charging process which makes them quite unstable against the interaction with the environment. However, by suitably engineering the environment fast self-discharging can be circumvented. In fact, single trapped ions are a well-known example of a realization of the quantum harmonic oscillator \cite{LB3}. Researchers have also harnessed these systems to showcase the development of artificially engineered reservoirs tailored for open quantum systems \cite{tm+0,RS+04}.  It may be possible to engineer the environment so that the presence of the environment slows down the battery self-discharging process or makes the battery relatively stable after switching off the laser or charging process \cite{CCF+20,XZ, SCC+19}. Keeping these future possibilities in mind, we conclude this paper with the hope that the present work will provide the seed for several future investigations (both theoretical and experimental) on QHO batteries.

\section*{Acknowledgment}
KG and AP acknowledge support from the QUEST scheme of the Interdisciplinary Cyber-Physical Systems (ICPS) program of the Department of Science and Technology (DST), India, Grant No.: DST/ICPS/QuST/Theme-1/2019/14 (Q80). KG also thanks Britant and Satish Kumar for fruitful discussions and suggestions.

\appendix
\section{Hamiltonian in Rotating Frame}
Consider that state $|\psi\rangle$ describes a system in stationary frame, and the same system is described as state $|\phi\rangle$  in a rotating frame which is rotating with angular frequency $\omega_f$.  These two states are related to each other by the equation
\begin{eqnarray}
    |\phi\rangle= U|\psi\rangle. \label{eq:A1}
\end{eqnarray}
Considering $\hbar=1$, and using Schrodinger equation and the identity $UU^\dagger=U^\dagger U=I$, we can easily obtain two alternative expressions for the Hamiltonian of the rotating frame ${\cal H_R}$ as follows
\begin{align}
 i\frac{d}{dt}|\phi\rangle &= i\frac{dU}{dt}|\psi\rangle + iU\frac{d}{dt}|\psi\rangle \nonumber\\ 
\implies i\frac{d}{dt}|\phi\rangle &= i\frac{dU}{dt}U^\dagger U|\psi\rangle + U{\cal H}U^\dagger U|\psi\rangle \nonumber\\
\implies {\cal H_R}|\phi\rangle &= (i\frac{dU}{dt}U^\dagger + U{\cal H}U^\dagger)|\phi\rangle \nonumber \\
\implies  {\cal H_R} &= i\frac{dU}{dt}U^\dagger + U{\cal H}U^\dagger \label{eq:A2}
\end{align}
Using the identity , we can express ${\cal H_R}$ in two equivalent but alternate forms as
\begin{align}
{\cal H_R} &= -iU\frac{dU^\dagger}{dt} + U{\cal H}U^\dagger, \label{eq:A3}
\end{align} 
and
\begin{align}
{\cal H_R} &= U^\dagger {\cal H}U - iU^\dagger\frac{dU}{dt}. \label{eq:A4}   
\end{align}
In our case, we may consider
\begin{align}
    U=e^{-i\omega_f(a^\dagger a + b^\dagger b)t}.\nonumber
\end{align}
Substituting this expression for $U$ and expression of ${\cal H}$ from Eq.\eqref{eq:1} into \eqref{eq:A2} we can obtain
\begin{eqnarray}
{\cal H_R} =  &\omega (a^\dagger a +  b^\dagger b) + g(ab^\dagger + a^\dagger b)\nonumber
\\
+& F(a +  a^\dagger)- \omega_f (a^\dagger a)-\omega_f (b^\dagger b). \nonumber 
\end{eqnarray}
Equivalently, we can express $\cal H_R$ as
\begin{eqnarray}
{\cal H_R} = & \delta (a^\dagger a +  b^\dagger b) + g(ab^\dagger + a^\dagger b) + F(a +  a^\dagger).
\end{eqnarray}
\\

\section{Battery Energy Expression}
From Eq.\eqref{eq:13} and Eq.\eqref{eq:14}, we can write
\begin{eqnarray}
\frac{d^2\langle b\rangle}{dt^2}+(2i\delta +\frac{\gamma_B}{2})\frac{d\langle b\rangle}{dt}-(\delta^2-g^2-i\frac{\gamma_B\delta}{2})\langle b \rangle=-gF \nonumber
\end{eqnarray}
This is a linear differential equation with constant coefficients. Solution of this differential equation can be obtained as
\begin{eqnarray}
    \langle b (\tau)\rangle = c_1 e^{m_1 \tau}+c_2 e^{m_2 \tau}+ \frac{gF}{(\delta^2-g^2-i\frac{\gamma_B \delta}{2})}, \label{eq:B1}
\end{eqnarray}
where $m_1=\frac{1}{2}\{-\frac{\gamma_B}{2}-2i(\delta-g)\}$, $m_2=\frac{1}{2}\{-\frac{\gamma_B}{2}-2i(\delta+g)\}$, and $c_1$ and $c_2$ are integral constants.
To find $c_1$ and $c_2$, we put initial conditions $\langle b \rangle(0)=0$, $\langle \dot{b}\rangle(0)=0$. These initial conditions lead to following relations

\begin{align}
c_1 + c_2 &= -\frac{gF}{\delta^2 - g^2 - i\frac{\gamma_B\delta}{2}} \label{eq:B2} \\
c_1 &= -\frac{c_2 m_2}{m_1} \label{eq:B3}
\end{align}

 Now, by using mean-field approximation we can write energy of the battery as
\begin{align}
E_B &\approx \omega_o \langle b^\dagger(t) \rangle \langle b(t) \rangle \nonumber \\
E_B &\approx c_1^*c_1 e^{t(m_1^*+m_1)} + c_2^*c_2 e^{t(m_2^*+m_2)}+ c_1^*c_2 e^{t(m_1^*+m_2)}
\nonumber \\
&\quad + c_1c_2^* e^{t(m_1+m_2^*)} + c_1^* e^{m_1^*t}\frac{gF}{\delta^2-g^2-i\frac{\gamma_B\delta}{2}} \nonumber \\
&\quad + c_1 e^{m_1t}\frac{gF}{\delta^2-g^2+i\frac{\gamma_B\delta}{2}}+ c_2^* e^{m_2^*t}\frac{gF}{\delta^2-g^2-i\frac{\gamma_B\delta}{2}} 
\nonumber \\
&\quad + c_2 e^{m_2t}\frac{gF}{\delta^2-g^2+i\frac{\gamma_B\delta}{2}}+\frac{g^2F^2}{(\delta^2-g^2)^2+\frac{\gamma_B^2\delta^2}{4}}. \nonumber \\ \label{eq:B4}
\end{align}
We solve \eqref{eq:B2} and \eqref{eq:B3} to get $c_1$ and $c_2$. We put $c_1$ and $c_2$ into \eqref{eq:B4} to obtain the energy expression of the battery.

\nocite{*}

\bibliography{apssamp}
\end{document}